\documentclass[prl,twocolumn,showpacs,preprintnumbers,amsmath,amssymb,superscriptaddress]{revtex4}
\usepackage{graphicx}
\usepackage{dcolumn}
\usepackage{bm}
\usepackage{amsmath}
\usepackage{amsfonts}
\usepackage{amssymb}
\usepackage{color}
\usepackage{epsfig,float,afterpage,amssymb,wrapfig,psfrag}
\usepackage[normalem]{ulem}

\begin{document}
\title{Reply to "Comment on 'Universal out-of-equilibrium transport in Kondo-correlated quantum dots'"}
\author{Enrique Mu\~{n}oz}
\affiliation{Facultad de F\'isica, Pontificia Universidad Cat\'olica de Chile, Casilla 306, Santiago 22, Chile.}
\author{C.~J.~Bolech}
\affiliation{Department of Physics, University of Cincinnati, Cincinnati, Ohio 45221-0011, United States.}
\author{Stefan Kirchner}
\affiliation{Max Planck Institute for the Physics of Complex Systems,
01187 Dresden, Germany.}
\affiliation{Max Planck Institute for Chemical Physics of Solids,
01187 Dresden, Germany.}



\maketitle

A recent comment~\citep{AAAcomment} on our work~\cite{Munoz.13} by A.A.Aligia claims that we "made mistakes in the evaluation of the lesser quantities $\Sigma^{-+}$ and $G^{-+}$" as we allegedly "neglected a term proportional to the non-interacting lesser Green's function $g^<$ in the expression for $G^<$". A.A.Aligia further claims that the distribution function of the single-particle selfenergy of the quantum dot in the Fermi liquid regime, {\it e.g.} at small bias voltage ($V$), low temperature ($T$), and small frequency ($\omega$), is continuous.
These claims are based on a comparison of the particle-hole symmetric case with results obtained from the RPTU approach of Refs.~\citep{Aligia.12,AAAcomment}. 
We disagree with these claims.
In fact, a comparison of our approach~\cite{Munoz.13}  with the numerical renormalization group (NRG) shows perfect agreement for the symmetric case~\cite{Merker.13}.
As we will show below, the discrepancy between ours  and the results of Ref.~\cite{Aligia.12}, alluded to in Ref.~\citep{AAAcomment},
can be traced back to a violation of certain Ward identities in
Refs.~\cite{AAAcomment,Aligia.12}.
In contrast, the approach of Ref.~\cite{Munoz.13} respects these Ward identities.

That $G^{-+}=G^a \Sigma^{-+} G^r$, Eq.(80) of~\cite{suppl} is indeed a correct form of the Dyson equation for the lesser component in  steady state is a standard result discussed {\it e.g.} in Refs.~\cite{Kamenev,Rammer,Wingreen.94}
and it is also shown in  Eqs.(71)-(80).
From the derivation in  Eqs.(71)-(80) it is obvious that  $\mathbf{g}$
is given in a basis of local, exact (un-dampened) eigenfunctions. 
The absence of the bare lesser propagator, a purely imaginary quantity, reflects the fact that no regularization of the bare propagator is necessary in the presence of selfenergy terms.  Physically, this states that the steady state is independent of the initial condition of the interacting region. This generally accepted statement can be made more rigorous in the present context~\citep{Doyon.06}.
Thus, the dot is initially, {\it i.e.} in the infinite past,  in an arbitrary, non-interacting local state disconnected from the leads and described by a propagator ${\mathbf g}$.
The hybridization is switched on and after waiting sufficiently long the system is described by a Green's function $\mathbf{g}_{\sigma\omega}^{(0)}$, then the local Coulomb interaction is added resulting in propagators $\mathbf{g}_{\sigma\omega}$ at particle-hole symmetry and $\mathbf{G}_{\sigma\omega}$ away from it.
Both ${\mathbf g}$ and $\mathbf{g}_{\sigma\omega}^{(0)}$ can serve as bare propagators in a Dyson equation for $\mathbf{G}_{\sigma\omega}$.
Thus, the term $2i\Delta$
that we allegedly missed  is not neglected but is part of the selfenergy $\Sigma$ in Eqs.(73)-(76).
It should also be clear from {\it e.g.} Eqs.(11)-(17) that the term in question is included in the Dyson equation as part of the Green's functions $\mathbf{g}_{\sigma\omega}^{(0)}$ and $\mathbf{g}_{\sigma\omega}$, as indicated by the presence of subindexes in our notation.
As a result, an identity of the steady state of the model ensues, Eq.(81) of~\cite{suppl} (see also Eqs. (74),(75)), relating the distribution function $F$ of the local Green's function to $\tilde{F}$, the distribution function of the associated selfenergy: 
$\tilde{F}=F$.
In the low $\omega$, low $V$, low $T$ regime we are concerned with we find no change to the distribution function from those obtained from $\mathbf{g}_{\sigma\omega}^{(0)}$ up to the order considered, see {\it e.g.} Eq.(50).
We note that a comparison of our approach~\cite{Munoz.13} in equilibrium, {\it i.e.} for vanishing bias voltage $V$,  with the NRG shows perfect agreement for the symmetric case~\cite{Merker.13}.
As the term $2i\Delta$ that we allegedly missed~\cite{AAAcomment} does not vanish when
$V=0$, its neglect would show up in equilibrium properties. Thus, Ref.~\cite{Merker.13} is further proof that the critique of the preceding comment is unwarranted.

Where do the differences shown in Fig.1 of~\cite{AAAcomment} come from?
The approach of Ref.~\citep{Aligia.12} is based on approximating the lesser component of the selfenergy matrix by a single diagram, Eq.(16) of Ref.~\citep{Aligia.12}. This spoils Ward identities. Our construction~\cite{Munoz.13,Scott.13}  is based on exact Ward identities for the symmetric model~\cite{Oguri.01}.
According to these identities, the first derivative in voltage is shown to be related to derivatives
of the corresponding selfenergy matrix in equilibrium (see Eq.(93) of Ref.~\cite{suppl} or Eq.~(8) of Ref.~\cite{Oguri.01}):
\begin{equation}
\left.\frac{\partial\mathbf{\Sigma}_{\sigma\omega}}{\partial(eV)}\right|_{V=0} = -\bar{\alpha}\left(\frac{\partial}{\partial\omega}
+\frac{\partial}{\partial E_{d}} \right)\mathbf{\Sigma}_{\sigma\omega}^{eq}\label{eq_Ward}.
\end{equation}
This derivative is proportional to $\bar{\alpha} = (\alpha_L\Gamma_L -
\alpha_R\Gamma_R)/(\Gamma_R + \Gamma_L)$. For $\bar{\alpha}=0$ (or $\Gamma_L=\Gamma_R$, $\alpha_L=\alpha_R$, the case considered in~\citep{AAAcomment} where $\bar{\alpha}$ is denoted $\gamma$),
 no linear-in-$V$ terms can thus be present
in any of the components of the non-equilibrium selfenergy matrix $\mathbf \Sigma_{\sigma \omega}$.
The imaginary part of the retarded and advanced selfenergy components of $\mathbf \Sigma_{\sigma \omega}$ in equilibrium ({\it i.e.} at $V=0$) do not contain any terms linear in $\omega$ or $T$ and only contains terms of the order $T^2$, $\omega^2$
and higher. Therefore,  in equilibrium and at $T=0$, neither component of the imaginary part of the selfenergy  matrix 
contains a linear-in-$\omega$ term. Performing a Sommerfeld expansion shows that there are no linear-in-$T$ terms either in the off-diagonal components of $\mathbf \Sigma_{\sigma \omega}$.
Any linear contribution in $\omega$ or $T$  to $\mathbf{\Sigma}_{\sigma\omega}$ therefore
has to enter at finite voltage. 
At the lowest order such terms have to be of the form
$\omega V ~\partial^2\Sigma^{-+}/\partial\omega\partial V|_{V=0,\omega=0}$ and $T V ~\partial^2\Sigma^{-+}/\partial T\partial V|_{V=0,\omega=0}$. 
The exact Ward identity Eq.~(\ref{eq_Ward}) implies that these terms exactly vanish for $\bar{\alpha}=0$.
Hence, the quantity shown in Fig.1 of \citep{AAAcomment} cannot contain  terms linear in $\omega$ or terms proportional to $VT$.
Yet, $\Sigma^{-+}$ (or $\Sigma^<$ in the notation of Ref.~\citep{AAAcomment}) obtained from the method of Ref.~\citep{AAAcomment,Aligia.12} clearly contains such terms, see Fig.~1 of \citep{AAAcomment}.
The violation of Eq.~(\ref{eq_Ward}) is most clearly seen by noticing the linear-in-$\omega$ and linear-in-$V$ term
in Eq.~(20) of~\citep{Aligia.12}: In the interval $-\alpha V<\omega<\alpha V$, Eq.~(20) of~\citep{Aligia.12} can be written
\begin{equation}
i\Sigma^{<}(\omega)=\pi (\tilde{\rho}_0(0))^3 \tilde{U}^2 \beta^3 [4 \omega^2+3V^2-6\omega V  ]
\end{equation}
(with $\alpha_L=\alpha_R=\alpha=1/2$ and $\beta_L=\beta_R=\beta$ corresponding to $\gamma=0$, {\it i.e.} the case considered in Ref.~\citep{AAAcomment}).
The presence of such linear-in-$\omega$ and linear-in-$V$ terms in the RPTU approach of Ref.~\cite{Aligia.12,AAAcomment} is at odds with the exact Ward identity, Eq.(1).\\
An additional problem with Refs.~\cite{Aligia.12,AAAcomment} is the amount of overcounting inherent in this approach. First, the propagators are dressed with a selfconsistent Hartree term which results
in an $n$ dependence (here, $n$ is the average dot occupation) of all propagators. This $n$ is formally given as an integral
over $G^{<}(\epsilon)$ which itself depends on $n$ and $V$, see Eq.(6) of Ref.~\cite{Aligia.12}. On top of this selfconsistent theory, the author of \citep{AAAcomment},\citep{Aligia.12} adds the renormalized perturbation theory (RPT) of A. Hewson~\cite{Hewson,Hewson.93,Hewson.01} by bringing in the wave function renormalization $z$ without adjusting the counter terms to the modified procedure ({\it i.e.} the already performed selfconsistent renormalization).
\newline
In summary, no term $\sim 2i \Delta$ has been neglected in the lesser selfenergy and $\tilde{F}=F$ (see Eq.(81) of \cite{Munoz.13}) is an exact property of the steady state of the model considered in~\cite{Munoz.13}.
In the Fermi liquid regime of small $\omega, T, V$ the distribution function $F$ and hence $\tilde{F}$ are discontinuous.
The discrepancies that the comment~\citep{AAAcomment} alludes to originate  from a violation of Ward identities by the method of Refs.~\cite{Aligia.12,AAAcomment} while the results of Ref.~\cite{Munoz.13} respect these identities.

\end{document}